\documentclass[useAMS,usenatbib]{mn2e}
\usepackage{epsfig}
\bibpunct{(}{)}{;}{a}{}{,}
\title[Evolution of bimodal accretion flows]{Evolution of bimodal accretion
  flows}
\author[J. Gracia et al.]{J. Gracia,$^1$\thanks{E-mail: J.Gracia@lsw.uni-heidelberg.de}
  J. Peitz,$^2$
  Ch. Keller$^3$
  and M. Camenzind$^1$\\
$^1$Landessternwarte Heidelberg-K\"onigstuhl, 69117 Heidelberg, Germany\\
$^2$Institut f\"ur Astronomie und Astrophysik, Abt. Computational Physics,\\
Universit\"at T\"ubingen, Auf der Morgenstelle 10, 72076 T\"ubingen, Germany \\
$^3$Institut f\"ur Theoretische Astrophysik, Universit\"at Heidelberg,
 Tiergartenstra\ss{}e 15, 69121 Heidelberg, Germany}
\begin{document}

\date{Received 2002/ Accepted 2002}

\pagerange{\pageref{firstpage}--\pageref{lastpage}} \pubyear{2002}

\maketitle 

\label{firstpage}
\begin{abstract} 
  We consider time-dependent models for rotating accretion flows onto black
  holes, where a transition takes place from an outer cooling-dominated disc
  to a radiatively inefficient flow in the inner region.  In order to allow
  for a transition of this type we solve the energy equation, both, for the
  gas and for the radiation field, including a radiative cooling flux and a
  turbulent convective heat flux directed along the negative entropy gradient.
  The transition region is found to be highly variable, and a corresponding
  variation is expected for the associated total luminosity.  In particular,
  rapid oscillations of the transition radius are present on a timescale
  comparable with the local Keplerian rotation time.  These oscillations are
  accompanied by a quasi-periodic modulation of the mass flux at the outer
  edge of the advection-dominated accretion flow (ADAF). We speculate about
  the relevance for the high-frequency QPO phenomenon.
\end{abstract}

\begin{keywords}
  accretion, accretion discs -- black hole physics -- convection --
  hydrodynamics -- instabilities.
\end{keywords}

%% paper main body %%
%misc 
\newcommand{\sJoesele}{Juan Jos\'e Gracia Calvo}
\newcommand{\Ref}[1]{(\ref{#1})}

%beautyfying
%\newcommand{\prop}{\sim}
\newcommand{\prop}{\propto}

% abbreviations 
\newcommand{\Mdot}{\mbox{$\dot M$}}
\newcommand{\mdot}{\mbox{$\dot m$}}
\newcommand{\Msun}{{\rm M_\odot}}
\newcommand{\Rg}{\mbox{$\rm \R_g$}}

%partial derivatives 
\renewcommand{\d}[2]{\frac{\partial #1}{\partial #2}}
\newcommand{\sd}[2]{\partial_{#2} #1}
\newcommand{\dt}[1]{\d{#1}{t}}
\newcommand{\dr}[1]{\d{#1}{\R}}
\newcommand{\sdt}[1]{\sd{#1}{t}}
\newcommand{\sdr}[1]{\sd{#1}{\R}}
\newcommand{\D}[2]{\frac{d #1}{d #2}}
\newcommand{\Dt}[1]{\D{#1}{t}}
\newcommand{\Dr}[1]{\D{#1}{\R}}

% vectors and tensors
\renewcommand{\vec}[1]{\mbox{\boldmath$#1$}}
\newcommand{\ten}[1]{{\mathsf #1}}
\newcommand{\Cal}[1]{{\mathcal #1}}

% special vectors 
\newcommand{\er}{\vec {\hat e_r}}

% tensor operators
\newcommand{\Div}{\vec \nabla \! \cdot \!}
\newcommand{\lDiv}{\! \cdot \! \vec \nabla}
\newcommand{\Grad}{\vec \nabla}
\newcommand{\otprod}{\! \otimes \!}           % outer tensor product
\newcommand{\itprod}{\! \cdot \!}             % inner tensor (dot) product

\renewcommand{\div}[1]{{\rm div} #1}

% model Parameters 
\newcommand{\rtr}{\mbox{$R_{\rm tr}$}}

% indices
\newcommand{\tr}{_{\rm tr}}             % at transition
\newcommand{\kep}{_{\rm K}}             % index for keplerian
\newcommand{\vsc}{_{\rm vsc}}           % index for viscosity
\newcommand{\trb}{_{\rm trb}}           % turbulent
\newcommand{\strb}{_{\rm T}}            % turbulent short form
\newcommand{\adv}{_{\rm adv}}           % advective
\newcommand{\rad}{_{\rm rad}}           % radiative
\newcommand{\diff}{_{\rm diff}}         % diffusive 
\newcommand{\esc}{_{\rm esc}}           % escaping 
\newcommand{\inp}{_{\rm src}}           % input 

% physical variables, abbr.
\newcommand{\tvsc}{\ten\tau}        % viscous stress tensor 

\newcommand{\vkep}{v\kep}           % keplerian orbital velocity
\newcommand{\den}{\rho}             % density 
\newcommand{\R}{R}                  % radial coord
\newcommand{\h}{H}                  % disk height
\newcommand{\vr}{v_{\R}}            % radial velocity
\newcommand{\Om}{\Omega}            % angular velocity 
\newcommand{\Omk}{\Omega\kep}       % angular velocity 
\newcommand{\sound}{c_{\rm s}}      % adiabatic sound speed
\newcommand{\isound}{c_{\rm s}}     % isothermal sound speed
\newcommand{\asound}{c_{\rm ad}}    % adiabatic sound speed
\newcommand{\p}{p}                  % pressure
\newcommand{\en}{e}                 % energy density
\newcommand{\spe}{\varepsilon}      % specific energy density
\newcommand{\entrop}{s}             % entropy 
\newcommand{\urad}{u}               % radiative energy density
\newcommand{\pr}{p_\R}              % radial momentum density 

\newcommand{\sden}{\Sigma}          % surface density
\newcommand{\sen}{E}                % surface energy density
\renewcommand{\sp}{P}               % surface pressure 
\newcommand{\surad}{U}              % surface radiative energy density
\newcommand{\spr}{P_\R}             % radial momentum surface density 

\newcommand{\Q}{Q}
\newcommand{\Qvsc}{\Q^+\vsc}         % viscous dissipation
\newcommand{\Qrad}{\Q^-\rad}         % radiative cooling
\newcommand{\Qadv}{\Q^-\adv}         % (entropy) advective cooling
\newcommand{\Qtrb}{\Q^+\trb}         % turbulent convective transport

\newcommand{\Qdiff}{\Q^-\diff}       % radial radiative diffusion
\newcommand{\Qesc}{\Q^-\esc}         % vertical irradiation
\newcommand{\Qinp}{\Q^+\inp}         % thermal emssion

\newcommand{\Pk}{P\kep}              % kepler period at outer radius
\newcommand{\Pktr}{P\kep^{\rm tr}}   % kepler period at transition radius
\newcommand{\dtosc}{\Delta t_{\rm osc}}  % oscilation time interval

%\newcommand{}{}

% % journals
 \newcommand{\apj}{ApJ}      % {The Astrophysical Journal}
 \newcommand{\apjs}{ApJS}    % {The Astrophysical Journal Supplement}
 \newcommand{\apjl}{ApJL}    % {The Astrophysical Journal Letters}
 \newcommand{\mnras}{MNRAS}  % {Monthly Notices Royal Astronomical Society}
 \newcommand{\pasj}{PASJ}    % {Publ. of the Astronomical Society of Japan}
 \newcommand{\aap}{A\&A}     % {Astronomy and Astrophysics}
 \newcommand{\nat}{Nature}   % {Nature}
% \newcommand{\physrep}{Physics Reports}
% \newcommand{\araa}{Annual Review of Astronomy and Astrophysics}

% papers
%\defcitealias{NY94}{NY}
%\defcitealias{NIA}{NIA}

\section{Introduction}

The standard model for an optically thick, Keplerian accretion disc
\citep{SS73} explains the continuous emission observed from a variety of
accretion sources both on the galactic and extragalactic scale.  This
so-called SSD model cannot explain, however, the hard spectral component
observed in most compact accreting systems, e.g., in X-ray binaries or in
AGNs. This component is generally fitted by a power-law and
ascribed to a thin, radiatively inefficient plasma.  Such conditions are met
in an advection dominated accretion flow \citep[ADAF, for a review
see][]{NMQ98} where the heat, that is liberated locally by a disc annulus due
to viscous dissipation, is advected with the radial bulk flow and eventually
disappears into the central black hole, without being emitted towards the
observer.

The presence of reflection components and the Fe K$\alpha$ line observed in
X-ray spectra of BHXB's \citep{Ebisawa96} strongly suggest that cold gas from
a disc and hot gas must coexist in a small volume. A possible configuration is
the radial transition from an outer SSD to an inner ADAF.  A bimodal disc
model along this line, with the transition radius as a free parameter and with
the ADAF and SSD parts treated as separate entities, allows to fit the spectra
of persistent X-ray sources \citep{NMY96, MMK97, E01}. Furthermore, this model
does also explain the different spectral states observed in some soft
X-ray transients by varying the location of the transition radius \citep{E98}. 
A number of these sources do also show high-frequency quasi-periodic
oscillations (QPO) in their fourier-power spectra \citep{NoLe98}. These
examples demonstrate the necessity for a time-dependent treatment.

In a consistent theory, the disc transition radius is obtained naturally as a
function of flow parameters and boundary conditions. For the stationary limit,
this issue was addressed by \citet{MK00}. If the transition region between the
inner ADAF and the outer SSD is sufficiently narrow (i.e., $\delta R/R \ll 1$),
specific connection conditions can be derived which in turn significantly
constrain the parameter space available for global solutions. The most
restrictive condition is that the ADAF must support a minimum outward energy
flux into the SSD, which is then liberated as radiation. The nature of this
energy flux is still a matter of debate, in particular within the
one-dimensional approximation \citep{ABI00}. \citet{Honma96} and \citet{MK00}
specifically considered energy transport by turbulent convection, and conclude
that the transition radius for stationary flows is then constrained to a
radial domain with boundaries depending on the model parameters. 
% \citet{HC00b}
% studied two-dimensional accretion flows and found no transition, albeit in a
% region of the parameter space where a transition is not expected from steady
% models of \citet{MK00}.  

\section{The disc model}

We investigate time-dependent transition models by solving the full
Navier-Stokes equations for a vertically integrated axisymmetric disc.
In the energy equation for the gas we include dissipation of a 
turbulent convective energy flux, and in addition we solve a radiative
transfer equation in the diffusion approximation. 
Our disc model thus generalises the stationary transition model 
of \citet{MK00} to the non-stationary case.

Assuming vertical hydrostatic equilibrium, the vertical pressure scale height
is $\h = \isound/\Omk$, with $\isound$ the isothermal sound speed and $\Omk$ 
the Keplerian angular velocity associated with the pseudo-Newtonian gravitational
potential for the space-time of a non-rotating black hole with mass $M$ \citep{PW80}. 
As a result of the vertical averaging procedure \citep[e.g.,][]{NY95a},
the thermodynamic variables which appear in the model equations are all
vertically integrated quantities, such as surface density $\sden = 2 \h \den$,
total pressure $\sp$, internal energy density $\sen$ and radiation
energy density $\surad$. 

The continuity equation and the momentum equations in radial and azimuthal
% poloidal and toroidal 
direction are 
\begin{equation}
  \label{eq:FScont}
  \sdt{\sden} +  \Div \sden \vec v  = 0
\end{equation}
\begin{equation}
  \label{eq:FSradialmom}
  \sdt{\sden \vr} + \Div \sden \vr \vec v = - \sdr \sp 
                                 + \sden (\Om^2 - \Omk^2) \R 
                                 + (\Div \ten \tau)_\R
\end{equation}
\begin{equation}
  \label{eq:FSangmom}
  \sdt \sden l + \Div \sden l \vec v = (\Div \tvsc)_\phi \; \R,
\end{equation}
where $R$ is the cylindrical radius and $l$ is the specific angular momentum.
Viscous transport is modeled as a Navier-Stokes stress, i.e., the vertically
integrated stress tensor $\tvsc = 2 \mu \ten \sigma$, with $\ten \sigma$ the
shear tensor and $\mu$ the vertically integrated dynamical viscosity.
We adopt the $\alpha$-viscosity law of \citet{SS73} as a model for
anomalous turbulent viscosity, $\nu = \alpha \isound \h$, which 
results in $\mu = \sden \nu = \sden \alpha \isound \h$.  

The thermal disc structure is governed by the magnitude of the various heating
and cooling terms which contribute to the energy equation for both the gas and 
radiation field. The gas energy equation is written as
\begin{equation}
  \label{eq:FSenergy}
    \sdt \sen + \Div \sen \vec v + \sp_{\rm g} \Div \vec v 
%%         = \sden T (\sdt \sden\entrop + \Div \sden\entrop \vec v) 
         = \Qvsc +\Qtrb - \Qrad. 
\end{equation}
The vertically integrated viscous heating rate is given by 
$\Qvsc = (\tvsc \lDiv) \itprod \vec v$, with $\tvsc$ as given above.

As the entropy gradient in ADAFs decreases outward, ADAFs are convectively
unstable \citep{Honma96}. Thus we additionally consider a bulk energy
transport due to a convective energy flux $\vec F\trb$ with the associated net
heating rate given by $\Qtrb = -\Div \vec F\trb$.  
% deposited in the gas due to the inflow of convective energy flux is $\Qtrb =
% - \Div \vec F\trb$. 
A natural assumption is to set $F\trb$ proportional to the local entropy
gradient, i.e.,
\begin{equation}
  \label{eq:3}
  \vec F\trb = - \kappa\strb \, \sden T \, \Grad s,  
\end{equation}
with $\kappa\strb$ an effective diffusion coefficient. Following
\citep{Honma96, MKNN00, MK00} we parametrise $\kappa\strb$ in analogy with the
$\alpha$-viscosity prescription as $\kappa\strb = \alpha\strb \isound \h$.  In
general, $\alpha\strb \le \alpha$ is expected, since convective turbulence
contributes to the viscosity, but not all sources of viscosity produce a bulk
energy transport \citep{RG92}.  The rate at which heat is advected with the
flow is $\Qadv = \sden T (\vec v \lDiv) s = \Div \sen \vec v + \sp_{\rm g} \Div \vec
v$. The radiative cooling rate is $\Qrad = 2 c \den \h \kappa_P (a_R T^4 -
\surad/2\h)$, with $\kappa_P$
% = 0.24 \times 10^{25} \den T^{-7/2}
the Planck-mean opacity for free-free absorption.

Finally, the radiation energy equation is
\begin{equation}
  \label{eq:FSradiation}
  \sdt \surad + \Qdiff = \Qinp - \Qesc, 
\end{equation}
where $\Qinp = \Qrad$. Sinks for the radiative energy density originate from
radiative losses through the upper and lower disk surfaces $\Qesc$ and from
radial diffusion $\Qdiff = \Div \vec F\rad$. Following \citet{MK00}, we write
\begin{equation}
  \label{eq:FSFrad}
  \vec F\rad = - \frac{cH}{3\den\h\kappa_R}\nabla_{\!\R} \surad \vec {\hat e_\R},
\end{equation}
where $\kappa_R$ is the Rosseland-mean opacity including electron scattering
and free-free absorption. The equations are closed by an equation of state for
the pressure, $\sp = \sp_{\rm g} + \sp\rad$. We apply $\sp_{\rm g} = (\gamma - 1) E$, with
$\gamma$ the ratio of specific heats, together with $\sp\rad = U/3$.

We discretise the equations on a non-uniform radial mesh, and use a time
semi-implicit, operator-split scheme to advance the solution forward in time.
The most restrictive constraint on the time-step in explicit schemes is due to
radiative and viscous terms as a result of the associated time-scales.  We
avoid this problem by treating these terms implicitly.  Implicit methods for
finite volume discretisation of our equations result in a large sparse
matrix, where only the neighbourhood of the diagonal is populated.  The
coupling between individual equations gives raise to a few off-diagonal
entries.  If this coupling is weak, e.g. if the time increment is small, these
off-diagonal terms may be neglected at low order and the resulting matrix can
be inverted very efficiently. In fact, we first solve each equation
independently, then update the dynamical variables and finally recompute the
global defects with the full system. The procedure is now repeated until
global convergence is reached.  This scheme was originally proposed by
\citet{HR98} and is naturally advantageous for moderately coupled systems.  If
the coupling is too strong, the efficiency in terms of achievable physical
simulation time over computational CPU-time is very low, and may even drop
below that of an explicit scheme.

In this letter we restrict attention to the model parameters $(\mdot_i,
\alpha, \alpha\strb)$, where $\mdot_i$ is the mass accretion rate of the
initial model in Eddington units. The black hole mass and the adiabatic index
are held fixed at $M=10 M_\odot$ and $\gamma = 1.5$ in all models.
As initial data we take a self-similar ADAF profile of \citet{NY94}
with the advection efficiency parameter $f$ a step-function, i.e., 
with $f$ dropping from unity (initial ADAF) to zero (initial SSD) at 
an arbitrarily chosen initial transition radius.
%The radiation energy density is initially zero. 
At the outer boundary, located at $2000 \Rg$ -- where $\Rg = GM/c^2$ is the
gravitational radius -- we fix the accretion rate to $\mdot_i$ and the specific
angular momentum to the local Keplerian value.  Vanishing gradient boundary
conditions are applied to the remaining variables.  At the inner boundary at
$2.2 \, \Rg$, we apply free outflow boundary conditions.

\section{A fiducial solution}

\begin{figure}
  \begin{center}
%    \resizebox{\hsize}{!}{\includegraphics[angle=-90]{fig1.ps}}
    \resizebox{\hsize}{!}{\includegraphics[angle=0]{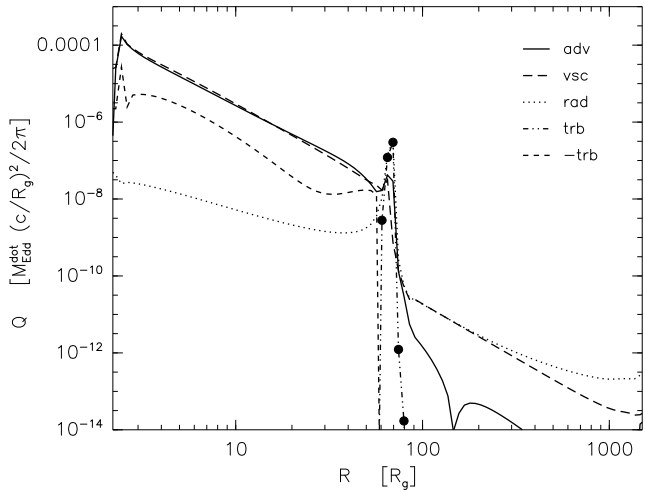}}
    \caption{    \label{fig:Qterms}
      Radial profile of the contributions to the gas energy balance (eq.
      \ref{eq:FSenergy}) for our fiducial solution. Note that $\Qtrb$ does act
      as a cooling or heating rate at different radii, i.e. it changes
      sign in the transition region. The dots superimposed on the $\Qtrb$
      curve indicate the spatial resolution.}
  \end{center}
\end{figure}

The transition radius $\rtr$ is defined as the location where the vertical
optical depth is equal to unity. Variables at the transition radius are
denoted by the subscript 'tr', and the subscript 'K' refers to Keplerian
motion.  Time is measured in units of the Keplerian orbital period taken at
the initial transition radius, $P\kep^{\rm tr}$, unless explicitly noted
otherwise.  Our unit of length is the gravitational radius $\Rg$.
%% $\Rg = GM/c^2$.
In Fig.~\ref{fig:Qterms} we plot the contributions to the energy balance for a
fiducial model with parameters $(\mdot_i, \alpha, \alpha\strb) = (0.021, 0.4,
0.4)$.  The transition in the initial conditions was located at $\rtr = 55
\Rg$.  These parameters were chosen to facilitate comparison with other
published results.

We generally identify three domains within the flow according to the magnitude
%balance 
of the different terms in the stationary energy equation (\ref{eq:FSenergy}),
even though our solutions are intrinsically time-dependent and cannot be
directly compared with steady state profiles.
% do not correspond to a true stationary state.
With increasing radial distance from the central black hole we identify the
inner ADAF where the optical depth is very low. Next we define the transition
region as the zone where the rotation is super-Keplerian.  Super-Keplerian
rotation is present whenever the transition region is narrow
\citep[cf.][]{AIL98}.  Between the outer edge of the transition region and the
outer edge of the disc we observe the SSD where the optical depth is very large.
%In our solutions the transition region is narrow for all times, and 
The radial profiles are never stationary, as discussed below.  Nevertheless,
the time-averaged profiles are rather similar to the stationary solutions of
\citet{MK00}.

%% Qterms und tau plot

\section{Evolution of the transition radius}

%% Location of transition radius and dtosc
\begin{figure*}
  \begin{center}
    \resizebox{.5\hsize}{!}{\includegraphics[angle=-90]{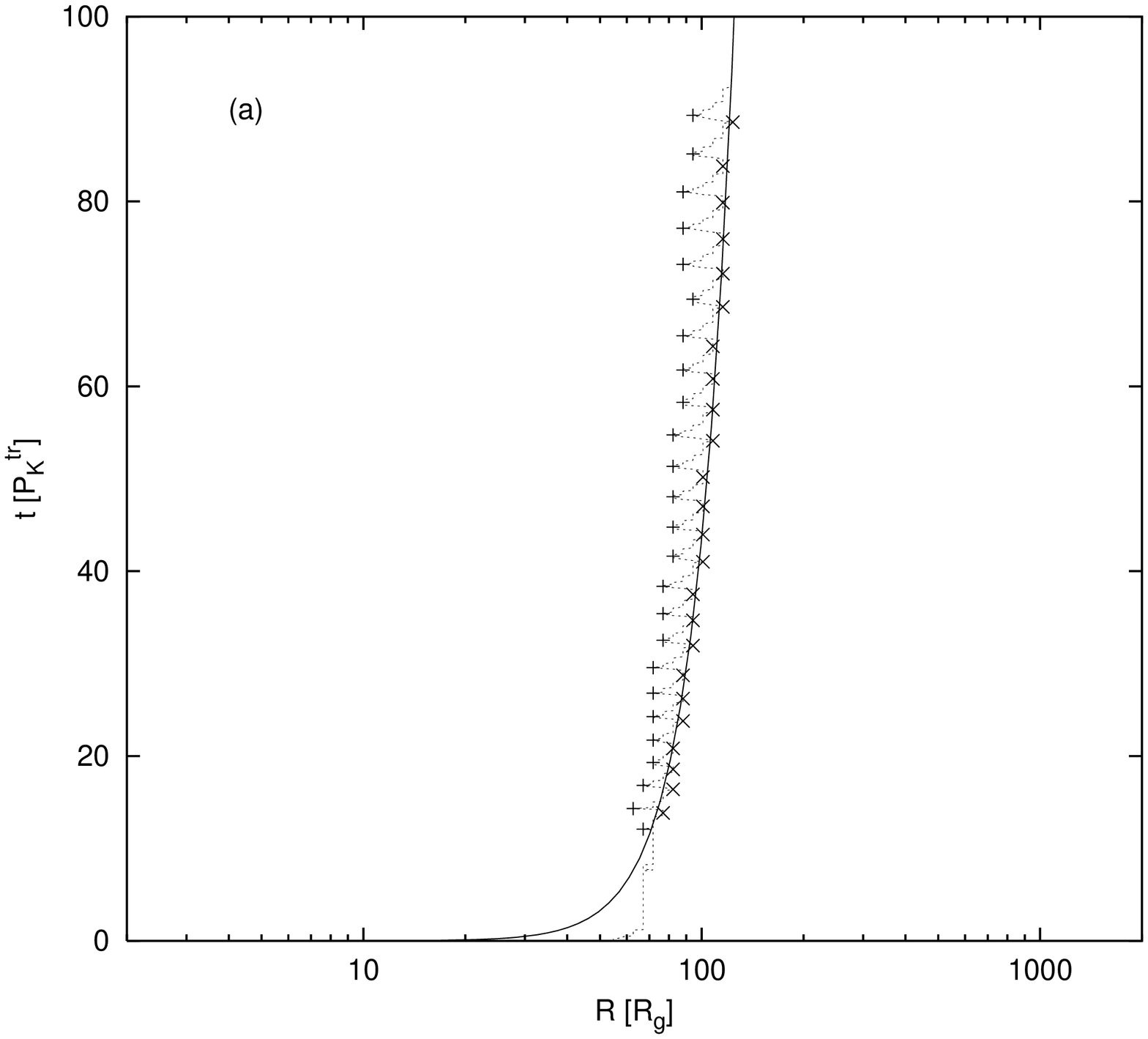}}%
    \resizebox{.5\hsize}{!}{\includegraphics[angle=-90]{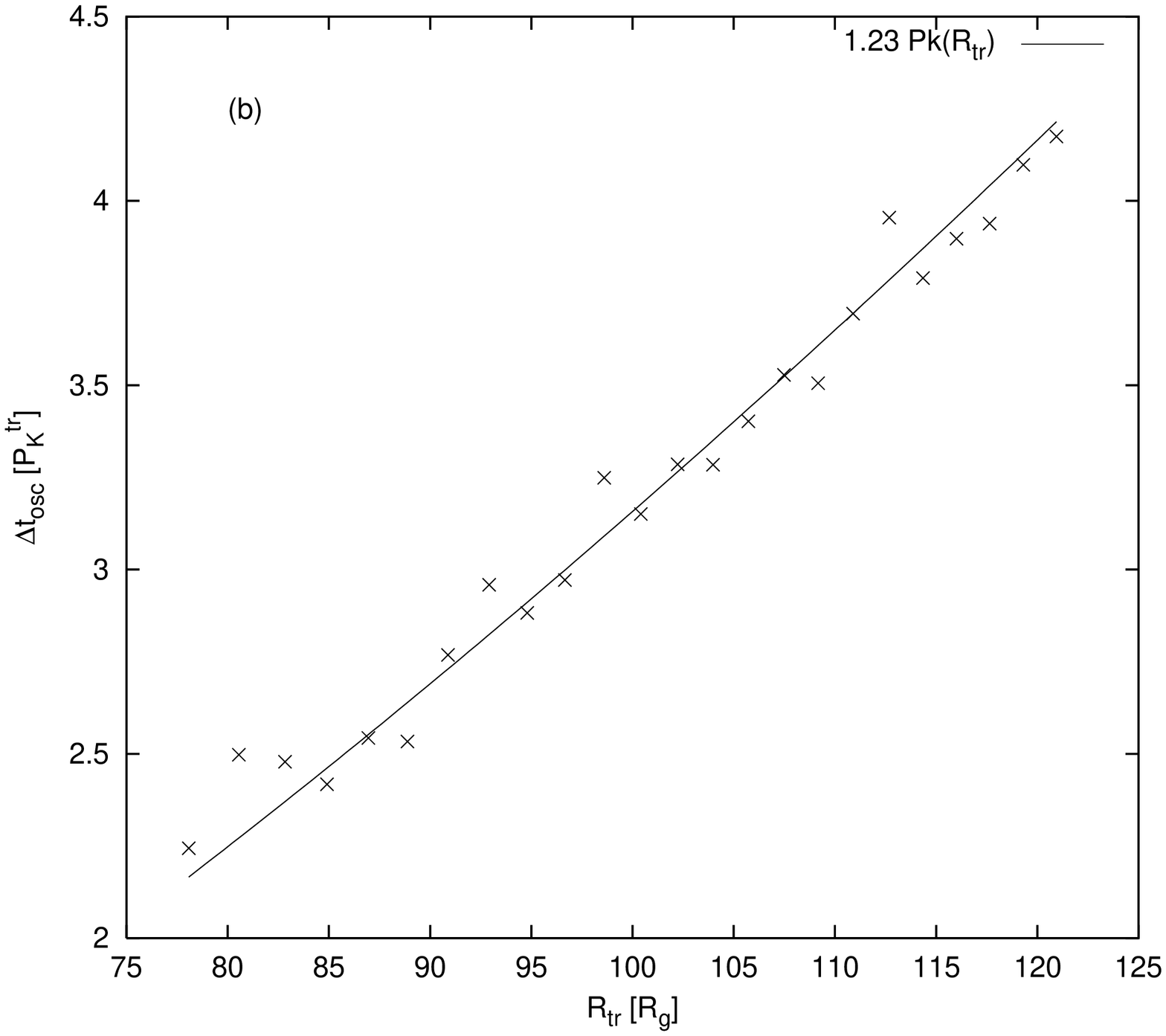}}

   \caption{    \label{fig:rtrloc} \label{fig:deltatosc}
     (a) temporal evolution of the transition radius $\rtr$ ({\em dotted}).
     The crosses $+$ and stars $\times$ indicate the innermost and outermost
     location of the transition radius. The latter is fitted by an analytical
     function $\Cal{R}_{\rm tr}(t)$ ({\em solid}).  (b) dependence of the
     oscillation cycle period $\dtosc$ on the transition radius $\Cal{R}_{\rm
       tr}$ for our fiducial model. In both panels time is measured in units
     of the Keplerian orbital period $P\kep^{\rm tr}$ at the initial
     transition radius located at $\rtr = 55 \, \Rg$.}
  \end{center}
\end{figure*}

The transition radius $\rtr$ derived from our solutions is a function of time.
We find that $\rtr$ is always located within the transition region.  After a
short initial relaxation phase, we generally observe a quasi-periodic time
variation of $\rtr$, superimposed on a much slower outward directed drift.
This behaviour is shown in Fig.~\ref{fig:rtrloc}(a).

The mass flow rate through the transition region is not constant in time, but
modulated with the same quasi-periodic pattern as $\rtr$. The transition
region tends to abruptly deposit cold gas into the ADAF. Whenever this
happens, the transition region grows at the expense of the ADAF.

On top of these rapid oscillations we observe a slow outward drift of the
transition radius and consequently, the SSD shrinks at this very rate while
the ADAF grows.  For the fiducial model introduced above, $\rtr$ grows from
initially $55 \, \Rg$ to finally $123 \, \Rg$ at $t = 92 \, P\kep^{\rm tr}$.
The associated drift velocity decreases with increasing $\rtr$ or,
accordingly, with increasing time.  A drifting behaviour of this type is
perceived in all our simulations.  However, the overall bimodal disc structure
is never disrupted and the transition region is always narrow in the sense
defined above.
% of \citet{AIL98}.
% !!  really deltaR/R?? Yes, if oscillation event is not considered !! 

% Our fiducial model did not converge
At $t = 92 \, P\kep^{\rm tr}$ the time-step of our simulation dropped
drastically to very small values, and the simulation was halted.
This behaviour was seen in all our simulations \citep{GPhD}.  Interestingly,
the values we measure for the transition radius at that time is nearly equal
to the maximum transition radius predicted for a stationary bimodal disc model
with the same parameters and boundary conditions \citep{MK00}. Near this maximum
transition radius the equations become very stiff, and thus the 
% i.e., the coupling between the different equations is very strong. In such situation the
% numerical method we currently employ is rather inefficient, thus the drastic
% drop of the convergence rate.
convergence of our numerical method decreases while the time-step freezes.
%Thus we argue that the break-down of our simulations is due to physical 
% rather than numerical grounds.  
The issue of the drift as such is further discussed in section
\ref{radial_drift}.

% beyond the convergent models there is no solution to the
% our system of equations for the parameters $(\mdot_i, \alpha, \alpha\strb,
% \rtr)$ we -- or in the case of $\rtr$ the flow itself -- chose.  This is an
% important issue and will be re-approached while discussing the radial drift of
% the transition radius in section \ref{radial_drift}.

\subsection{The oscillation cycle} \label{osc_event}

The period of an oscillation cycle $\dtosc$ increases with time.  In
Fig.~\ref{fig:rtrloc}(a) we plot the innermost location and the outermost
location which the transition radius takes during the complete evolution of
our fiducial model.  The outer location is treated as the generic transition
radius and fitted by an analytical function $\Cal{R}_{\rm tr}(t)$.  The
innermost location was in turn considered as the perturbed transition radius
in the course of the oscillation cycle and used to derive the oscillation
period $\dtosc$.

As shown in Fig.~\ref{fig:deltatosc}(b), a strong correlation is found between
$\dtosc$ and the generic transition radius $\Cal{R}_{\rm tr}(t)$. Furthermore,
$\dtosc$ is very close to (within a factor of order unity) the Keplerian
orbital period at the generic transition radius, $P\kep(\Cal{R}_{\rm tr}(t))$,
independent of model parameters $(\mdot_i, \alpha, \alpha\strb)$. However,
the width of the transition region is a function of the adiabatic
index $\gamma$ \citep{MK00} and therefore we expect it to modify the
proportionality factor between $\dtosc$ and $P\kep(\Cal{\R}_{\rm tr}(t))$.
%We find  $1/\omega_0 = 1.23 \pm 0.03$  

% A necessary condition for a narrow transition region is super-Keplerian
% rotation within the transition region, $l > l\kep$, i.e., in order to smoothly
% match the Keplerian rotation profile of the SSD, the specific angular momentum
% $l$ must decrease outward \citep{AIL98}. However, if $l$ decreases with
% increasing radius, $\sdr l < 0$, the flow is Rayleigh unstable. Consequently,
% a narrow ADAF-SSD transition region is the location of putative instability.
% % !! stimmt das?? Warum hat Abramowicz das nicht remarkt ??!!

A necessary condition for a narrow transition region is super-Keplerian
rotation within the transition region, i.e. $l > l\kep$ \citep{AIL98}.  In our
calculations the specific angular momentum profile decreases outward and
smoothly matches the Keplerian rotation profile of the SSD. However, if $l$
decreases with increasing radius, $\sdr l < 0$, the flow is Rayleigh unstable.
Consequently, a narrow ADAF-SSD transition region is the location of putative
instability.

Outward decreasing specific angular momentum profiles have previously been
obtained by e.g. \citet{Honma96} and \citet{MK00} as seen from their figures.
The specific angular momentum profile of \citet{AIL98} does not decrease
outward.  They used an adiabatic equation of state, which yields broader
transition regions, i.e.  $l$ changes slower near the outer edge of the
transition region and matches the SSD profile without actually decreasing.

Local stability analysis of a rotating flow predicts linear perturbations to
oscillate with frequency $\omega$ that is bounded by the epicyclic frequency
$\kappa$. Perturbations with $\omega^2 \approx \kappa_{\rm max}^2$ will
preferably grow or decay \citep[][for a review]{KM00, K01}.  If $\kappa^2$ is
negative the flow is dynamically unstable.  If $\kappa^2$ is positive the
perturbation will oscillate with frequency $\omega^2$ near $\kappa_{\rm
  max}^2$.  \citet{KM00} further argue that the oscillations will be trapped
within the transition region, since they are reflected at the transition
region boundaries.

In our solutions we find an upper limit $\kappa^2_{\rm max}/\Omk^2 \approx 2$,
i.e., the oscillation period $\dtosc/P\kep^{\rm tr}$ has a lower limit
$1/\sqrt{2}$ of order unity, consistent with our numerical results and with
our proposition that the oscillations have a physical nature.  On the other
hand, we note that $\kappa^2_{\rm max}$ sensitively depends on the angular
momentum gradient in the transition region, which is rather poorly resolved in
our current models.  Therefore, we cannot conclude from our simulations that
$\dtosc$ is always bounded by the above limit.

%% close-up oscillation event
\begin{figure*}
  \begin{center}
    \epsfig{file=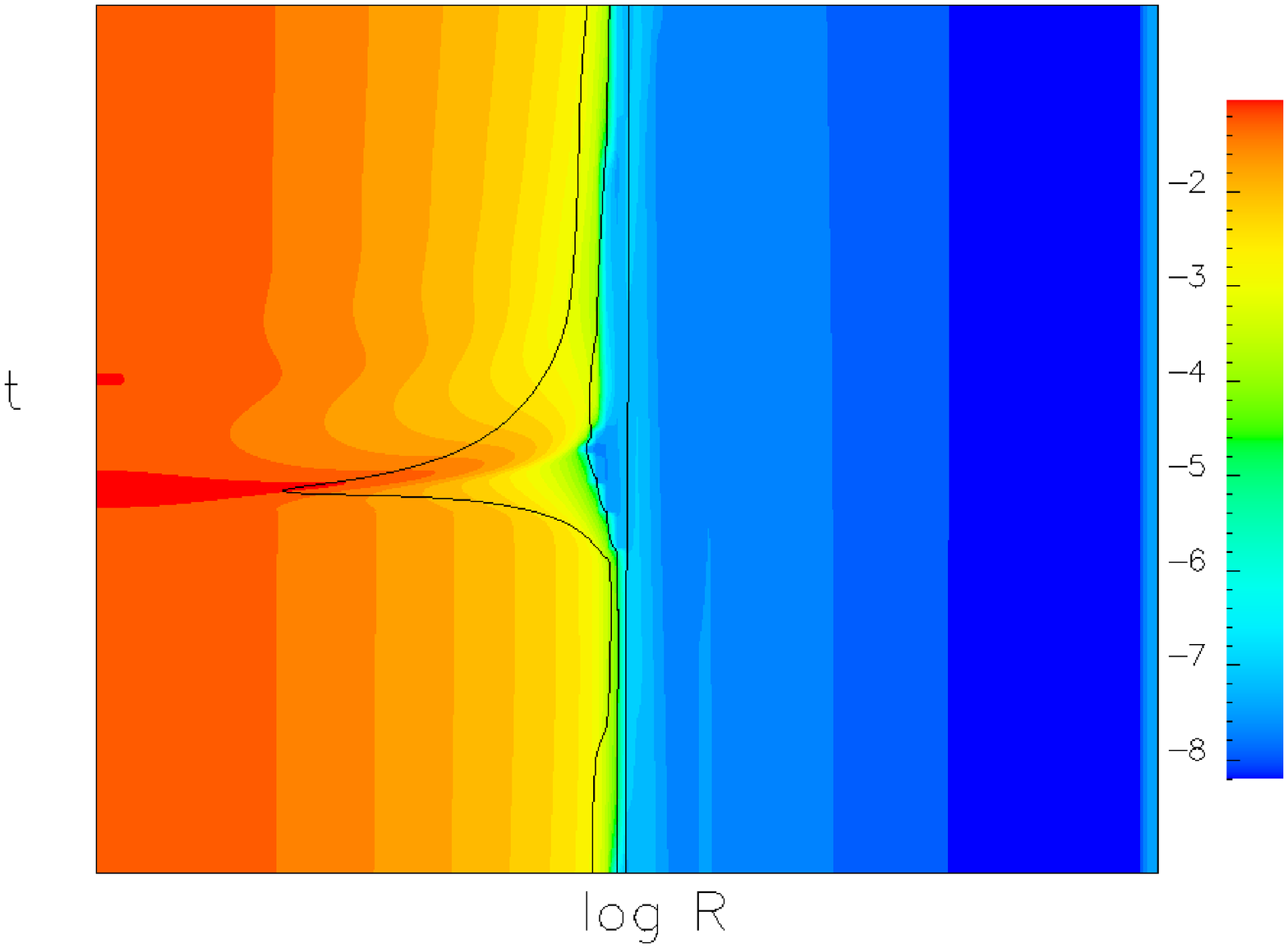, angle=-90, width=.5\hsize}%
    \epsfig{file=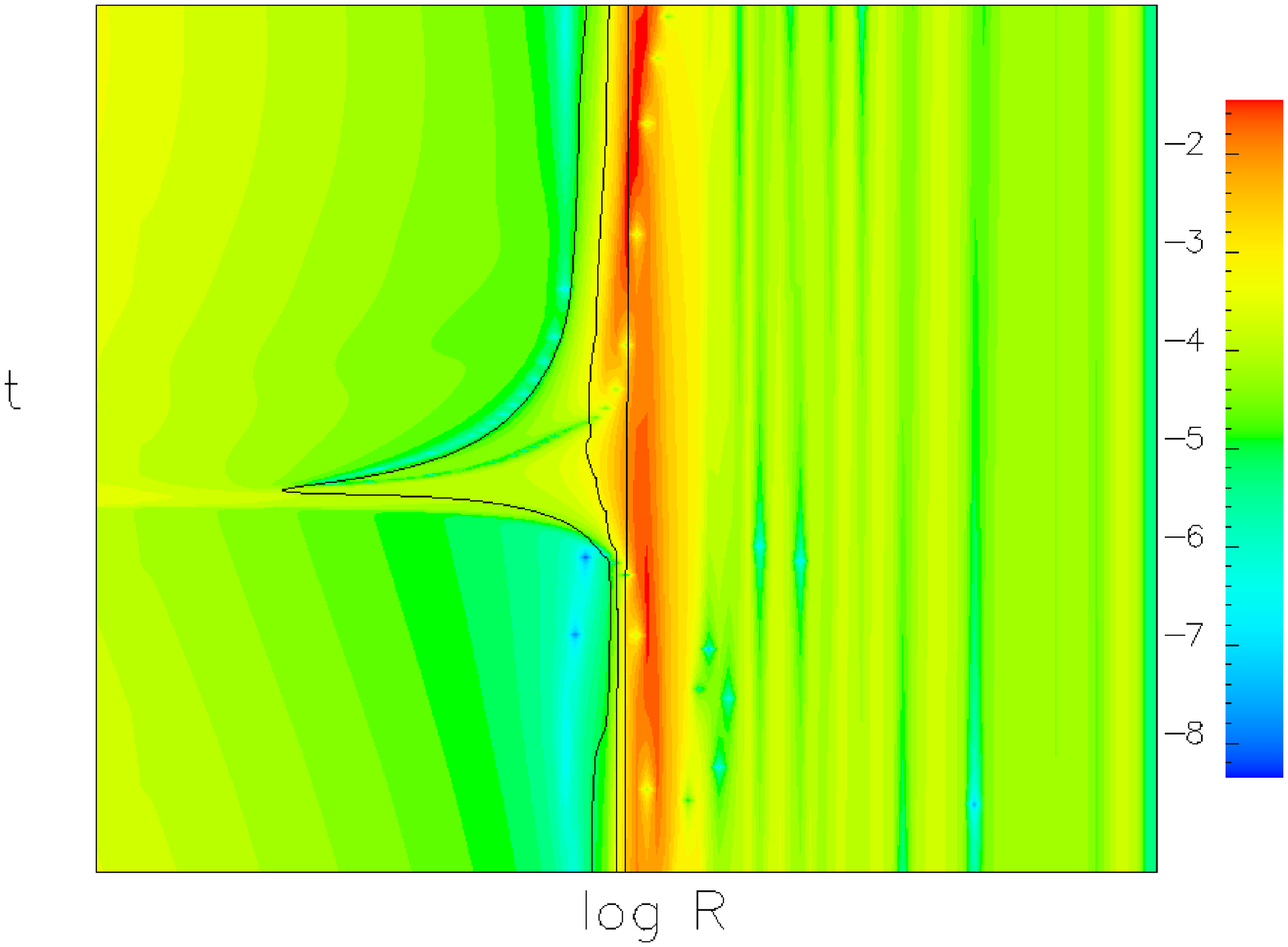, angle=-90, width=.5\hsize}
    \caption{    \label{fig:osc}
      Dynamics of the oscillations in the ADAF-SSD transition region.  On each
      panel radius $\R = (2\Rg \cdots 2000\Rg)$ increases from left to right
      on a logarithmic scale, while time $t = (0\cdots 0.92\,\Pktr(62 \,
      \Rg))$ runs from bottom to top on a linear scale. The individual panels
      show logarithm of temperature $\log T$ (left), and logarithm of
      vertically integrated radial momentum density $\log (|\sden \vr|)$
      (right). $T$ is measured in units of $(\gamma - 1) c^2 \mu m_{\rm
        p}/k_{\rm B}$ and $\sden \vr$ in units of $\Mdot_{\rm Edd}/(2\pi\Rg)$.
      The outer black contours indicate the zone of super-Keplerian rotation,
      i.e.  the transition region. The inner one, the transition radius
      $\rtr$.}
  \end{center}
\end{figure*}

We mentioned earlier that perturbations generated within the transition region
are expected to be effectively trapped within the same. In reality, this
simple picture does not survive the confrontation with numerical experiments
entirely unharmed. Perturbations from within the transition region do
back-react with the flow outside. Fig.~\ref{fig:osc} illustrates the dynamical
evolution of the flow during an oscillation cycle at high temporal resolution.

The oscillation cycle appears to be initiated by a perturbation as seen, e.g.,
in the radial momentum flux density, approaching the inner boundary of the
transition region.  The perturbation leaks into the ADAF, cold matter with
high angular momentum detaches from the transition region and penetrates the
ADAF as seen in Fig.~\ref{fig:osc}.  The ejected blob carries nearly constant
angular momentum, such that the region of super-Keplerian rotation, i.e., the
transition region, grows far into the ADAF.  When the perturbation reaches the
sonic radius, it is partly reflected as an outgoing wave and sweeps up hot
plasma towards the transition.  In continuation the ADAF relaxes slowly back
into the unperturbed state.  This causes high variability not only of the
transition region, but also of the inner ADAF, where most of the X-ray flux is
expected to originate. Hence the evolution of our models is strongly affected
by the existence of oscillations in the transition region.

\subsection{Radial drift or the strong ADAF principle} \label{radial_drift}

The outward drift of the transition region which we see in our simulations
does also require interpretation. It is well known that, for a given accretion
rate, the radial extent of the ADAF (if existent) is restricted by an upper
limit $\R_\tau$ \citep[e.g.][]{Chen95, ACKLR95}.  At every radius $\R >
\R_\tau$, the optical depth exceeds unity and radiative cooling is efficient,
such that the ADAF solution branch is no longer available.  For certain $\R <
\R_\tau$, the ADAF and SSD-like solutions are both available in principle.
From entropy considerations we expect, that nature realizes the ADAF whenever
possible, a proposition referred to as the {\em strong ADAF principle}
\citep{NY95b}.

Our models do indeed seem to support this hypothesis. The transition radius
$\Cal{R}_{\rm tr}(t)$ does always drift outward, thereby increasing the total
entropy of the entire solution. The drift is stopped once $\rtr$ has reached
the outermost location, $R\trb$, that is allowed from stationary analysis.
Note that $\R\trb$ is smaller than $\R_\tau$
% and given by the efficiency of turbulent convective energy transfer \citep{MK00}.
because turbulent convective energy transport is taken into account in our
models.  

For a given set of parameters, the specific entropy of the ADAF is, at any
radius of interest, larger than the specific entropy of the SSD. The
particular form applied for turbulent convective energy transport (namely along
the negative entropy gradient), implies that the transition region is always
supplied with energy from the inner ADAF.  As long as this energy surplus
cannot be completely shed by radiative losses, the transition region is heated
to nearly the local virial temperature and the flow switches to the ADAF mode.
As a consequence, the SSD is slowly evaporated radially at its inner edge.
This process does not stop until radiative energy shedding in the transition
region completely dominates energy supply by convection -- exactly the
definition of the outermost allowed transition radius $\R\trb$. Our models do
indeed evolve in general within the parameter space of \citet[see their Figs.
8, 9 and 10]{MK00}.

% If this picture is correct and the specific entropy gradient is indeed the
% driving force behind the radial drift of the transition region, then one would
% expect that the rate at which the transition radius $\rtr$ changes, is
% proportional to the entropy gradient at $\rtr$, i.e.
% \begin{equation}
% \label{eq:discRtrS}
%        %\frac{\mathrm{d^2}}{\mathrm{d}t^2}\rtr(t) \prop \d{s}{\rtr}
%                  \ddot{\R}_{tr}(t) \prop \d{s}{\rtr}.  !!??!!
% \end{equation}
% This is currently investigated and will be published in a future work. % \citep{GPC02}.

% In this context it would also be interesting to investigate if the flow
% properties down-stream of the transition could be related to properties
% up-stream of the transition by means of conserved quantities, at least in the
% steady limit. One of them is certainly the mass accretion rate $\mdot_i$.
% Another is possibly the total energy including contributions of the radiation
% field or some conversed energy flux. Such relations would in some sense
% correspond to generalised Rankine-Hugoniot relations. One could start to look
% for such relations at the outermost allowed location for the transition where
% the structure of energy budget and force balance in the transition is less
% complicated.

\section{Concluding remarks}

We have modelled the time evolution of bimodal accretion discs where an outer
cooling dominated disc matches an inner ADAF-type flow through a narrow
transition region. Our models show an intrinsically time-dependent behaviour
that can be understood from the dynamical instabilities predicted for
stationary ADAF-SSD transitions.  Specifically, oscillations with
frequency close to $\Omega\kep$ are excited in the transition region.
We associate these oscillations with a mutual local instability that 
originates from the radially decreasing specific angular momentum profile,
a characteristic feature of any narrow-transition bimodal disc.

We observe that the waves excited in the transition region leak into the inner
ADAF, but do not essentially propagate outward into the SSD. These
oscillations give rise to rapid periodic modulations of the ADAF and
the transition region. We estimated the total disk luminosity by
integrating the radiative cooling rate over the disk surface and again find a
periodic modulation with a variability of a few percent. This could possibly
explain the QPOs observed in X-Ray transient systems \citep[e.g.][and
references therein]{NoLe98}. \citet{KM00} estimated the oscillation frequency
in bimodal disc transitions to a small fraction $\sim 10^{-3}$ of
$\Omega\kep$, albeit under the assumption of small frequency and using a
rather crude estimate for the disc structure. Our numerical results
show frequencies very close to $\Omega\kep$ at the transition radius.

% At present, our models seem to over-estimate the oscillation frequency 
% observed in these sources. This is most likely due to the limited 
% spatial resolution and possibly due to the one-dimensional modelling.

We furthermore observe that the initial transition radius does slowly drift to
larger radii, i.e., even though a stationary transition is allowed for a wide
range of radii, the disc appears to favour the outermost possible transition
radius. This fits well into the {\em strong ADAF principle} hypothesis
\citep{NY95b}. Within our modelling, convective energy transport is the key to
make this drift possible, i.e., to evaporate the SSD radially until the
transition reaches its outermost allowed location. But even if we neglect
convective energy transfer as modelled in this paper, the large entropy
gradient across the transition still persists.
Significant transfer of heat from the ADAF to the SSD is a necessary condition
for the existence of a stationary transition front. Whether turbulence is the
relevant physical mechanism for this heat flux and whether the shape of the 
transition front is actually close to cylindrical cannot be addressed
by 1-dimensional modeling. The resolution of this issue does require at least
2-dimensional simulations in the parameter range that allows for a transition
in the 1-dimensional stationary case. This region of the parameter space, however, 
has not been addressed in a 2-dimensional simulation by Hujeirat \& Camenzind (2000).
The gap between vertical disc evaporation models \citep{MMH94, MLMH00} and the 
present 1-dimensional radial transition physics therefore remains to be closed
in a future work.

%% paper aftermath %%

\section*{Acknowledgements}
Part of this work was supported by the \emph{Deut\-sche
  For\-schungs\-ge\-mein\-schaft, DFG\/} (Son\-der\-for\-schungs\-be\-reich
439 and Sonderforschungsbereich 359).

\bibliographystyle{mn2e}
\bibliography{references}         % dont forget to submit the .bbl file, not .bib

\label{lastpage}
\end{document}